\documentclass[prl,aps,twocolumn,showpacs,floatfix]{revtex4}
\usepackage{subfigure}   
\usepackage{graphicx}  
\usepackage{dcolumn}   
\usepackage{amsmath}
\usepackage{amssymb}
\usepackage{color}

\usepackage{bm}
\usepackage{natbib}

\begin{document}
\bibliographystyle{apsrev}
\title{\bf Charge Localization Dynamics induced by Oxygen Vacancies \\
on the Titania TiO$_2$(110) Surface}

\author{Piotr M. Kowalski} 
\altaffiliation[Present address: ]{Helmholtz Centre Potsdam, Telegrafenberg, 14473 Potsdam, Germany}
\affiliation{Lehrstuhl f\"ur Theoretische Chemie, Ruhr--Universit\"at Bochum,
44780 Bochum, Germany}

\author{Matteo Farnesi Camellone} 
\affiliation{Lehrstuhl f\"ur Theoretische Chemie, Ruhr--Universit\"at Bochum,
44780 Bochum, Germany}

\author{Nisanth N. Nair} 
\altaffiliation[Present address: ]{Department of Chemistry, Indian Institute of Technology Kanpur, Kanpur 208016, India}
\affiliation{Lehrstuhl f\"ur Theoretische Chemie, Ruhr--Universit\"at Bochum,
44780 Bochum, Germany}

\author{Bernd Meyer}
\altaffiliation[Present address: ]{Interdisziplin\"ares Zentrum f\"ur Molekulare
  Materialien (ICMM) and Computer-Chemie-Cen\-trum (CCC), Universit\"at
  Erlangen-N\"urnberg, 91052 Erlangen, Germany}
\affiliation{Lehrstuhl f\"ur Theoretische Chemie, Ruhr--Universit\"at Bochum,
44780 Bochum, Germany}

\author{Dominik Marx}
\affiliation{Lehrstuhl f\"ur Theoretische Chemie, Ruhr--Universit\"at Bochum,
44780 Bochum, Germany}

\date{\today}
\begin{abstract}
The dynamics of an F--center created by an oxygen vacancy
on the $\mathrm{TiO_{2}(110)}$ rutile surface 
has been investigated using {\it ab initio} molecular dynamics.
These simulations uncover a truly complex, time-dependent behavior of 
fluctuating electron localization topologies in the vicinity of the oxygen vacancy. 
Although the two excess electrons are found to
populate preferentially
the second subsurface layer, they occasionally visit
surface sites and also the third subsurface layer.
This dynamical 
behavior of the excess charge
explains 
hitherto conflicting interpretations
of both theoretical findings and experimental data. 
\end{abstract}


\maketitle

Titanium dioxide ($\mathrm{TiO_{2}}$) is one of the most thoroughly investigated  
metal oxides, due to its broad range of uses in several key technologies including  
heterogeneous catalysis, 
pigment materials, 
photocatalysis,
and energy production, 
to name but a few~\cite{oregan,hammergold,imagawa}. 
It is well known that  
bulk and surface defects 
govern the 
properties of titania,
and are thus of fundamental importance in virtually all its applications~\cite{bennet,pang,lamb}. 	 
The most common point defects on the $\mathrm{TiO_{2}(110)}$ rutile surface are oxygen vacancies 
($\mathrm{O}_{\rm v}$) 
in the two--fold coordinated O~rows and $\mathrm{Ti}$
interstitials~\cite{Yim,WS08}. 
In particular, 
removal of an $\mathrm{O}$ atom gives rise to two excess electrons
and the appearance of new 
electronic states in the band gap at
about 0.7--0.9~eV below the conduction
band edge creating an F--center~\cite{henrich,diebold,sauer}.
Although the two excess electrons can in principle 
be localized
on any $\mathrm{Ti}$
atom,
they 
are believed
to preferentially occupy
specific $\mathrm{Ti}$--3${\it d}$ orbitals, thus formally  
creating $\mathrm{Ti^{3+}}$ sites~\cite{diebold,selloni}.
In stark contrast,
recent experiments~\cite{morgante}
suggest a qualitatively different viewpoint:
charge localization is found to be more 
disperse, with the excess electrons being shared by  
several surface and subsurface $\mathrm{Ti}$
ions.
Furthermore, 
STM/STS experiments have revealed
charge delocalization involving more
than ten $\mathrm{Ti}$ sites~\cite{minato}.

Unfortunately, 
different computational methods yield conflicting results~\cite{sauer}.
Local/semilocal density functionals (LDA/GGA) predict a rather delocalized defect level 
for O~vacancies on $\mathrm{TiO}_{2}(110)$ with an energy right at the bottom 
of the conduction band~\cite{sauer}.
However, it is well known that such functionals bias against localization on
strongly correlated 
{\em d}--states, and hence alternative methodologies are welcome.
Recent studies of defective TiO$_2$ 
surfaces~\cite{divalentin,thornton,pacchioni,morgan,morgan1,filippone,CHS08} 
have focused on ``pragmatic and practical'' correction schemes 
using hybrid functionals or 
a Hubbard correction.  
Although both 
schemes
yield the expected gap states, 
they each predict vastly different localization topologies of the excess charge.

\begin{figure}
  \begin{center}
    \begin{tabular}{cc}
       \subfigure[]{\includegraphics[scale=0.1,angle=0]{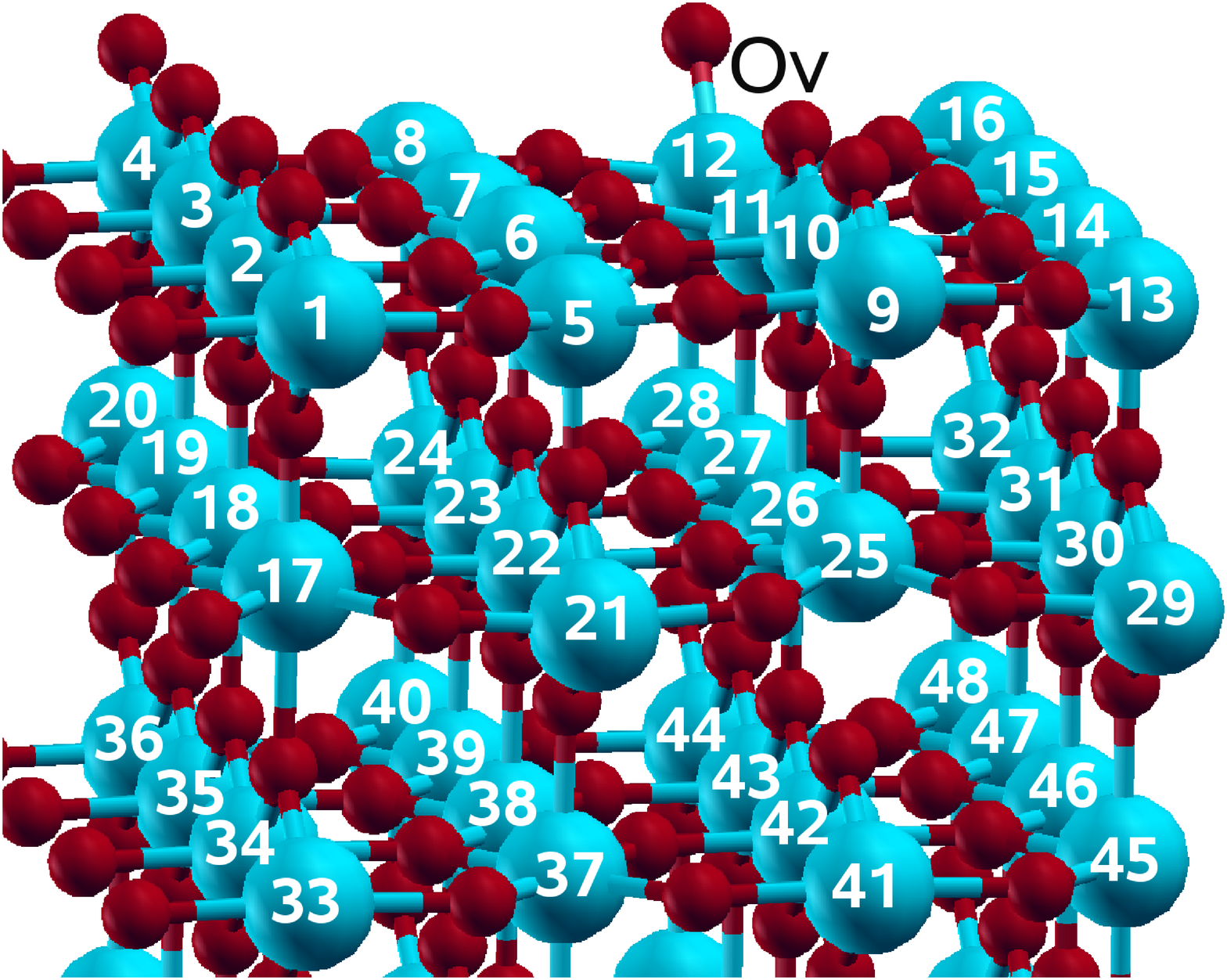}} &
       \subfigure[]{\includegraphics[scale=0.12,angle=0]{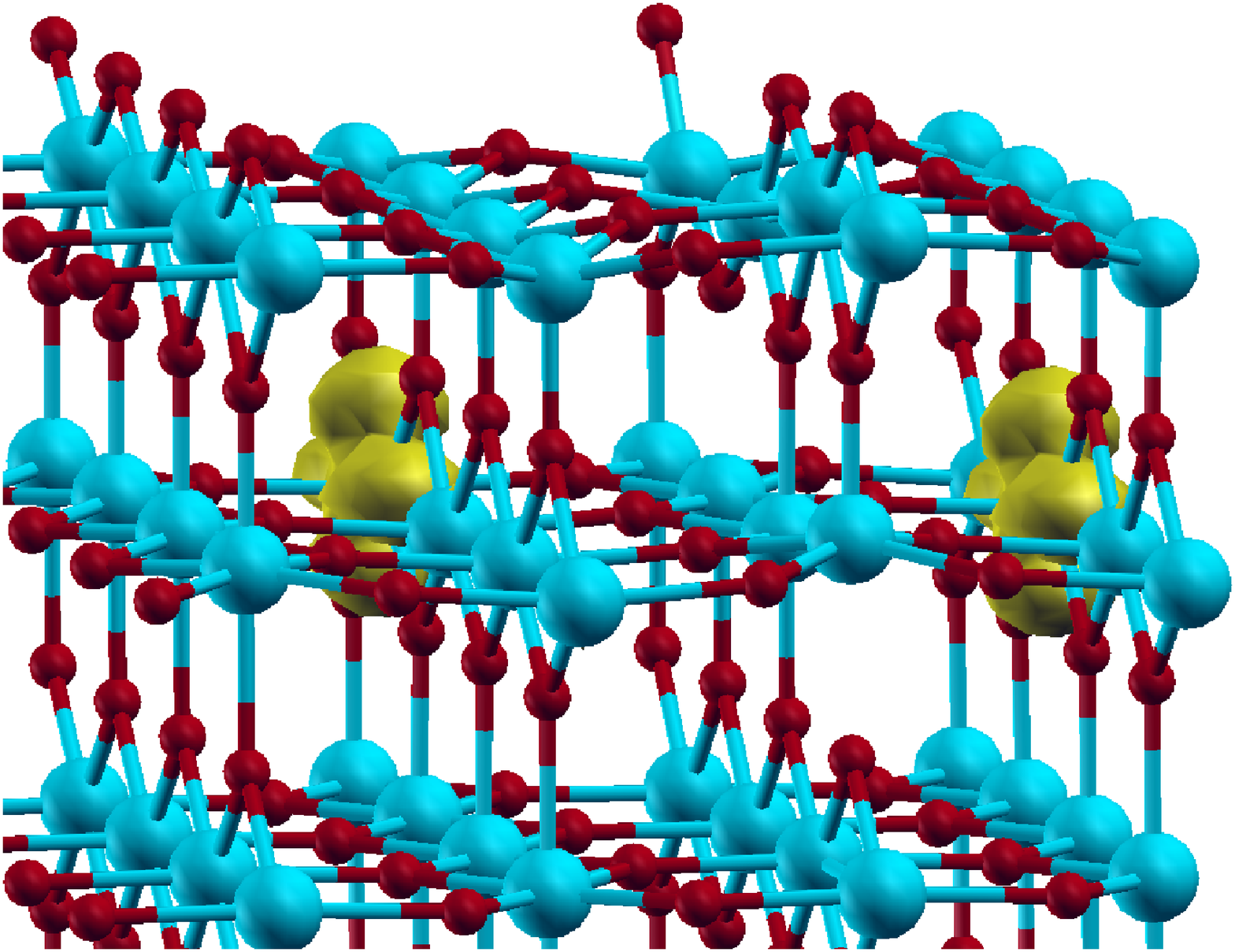}} \\
       \subfigure[]{\includegraphics[scale=0.12,angle=0]{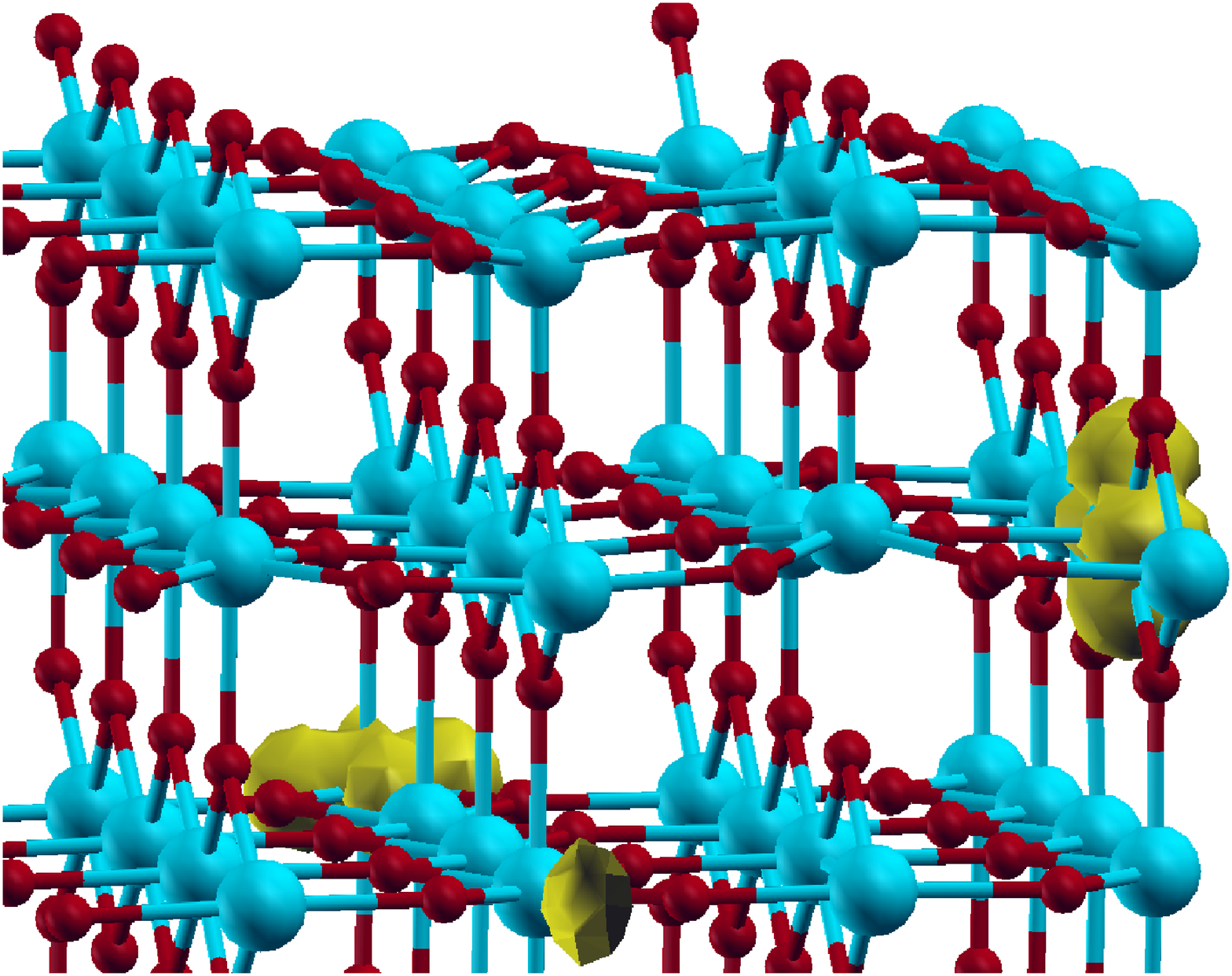}} &
       \subfigure[]{\includegraphics[scale=0.12,angle=0]{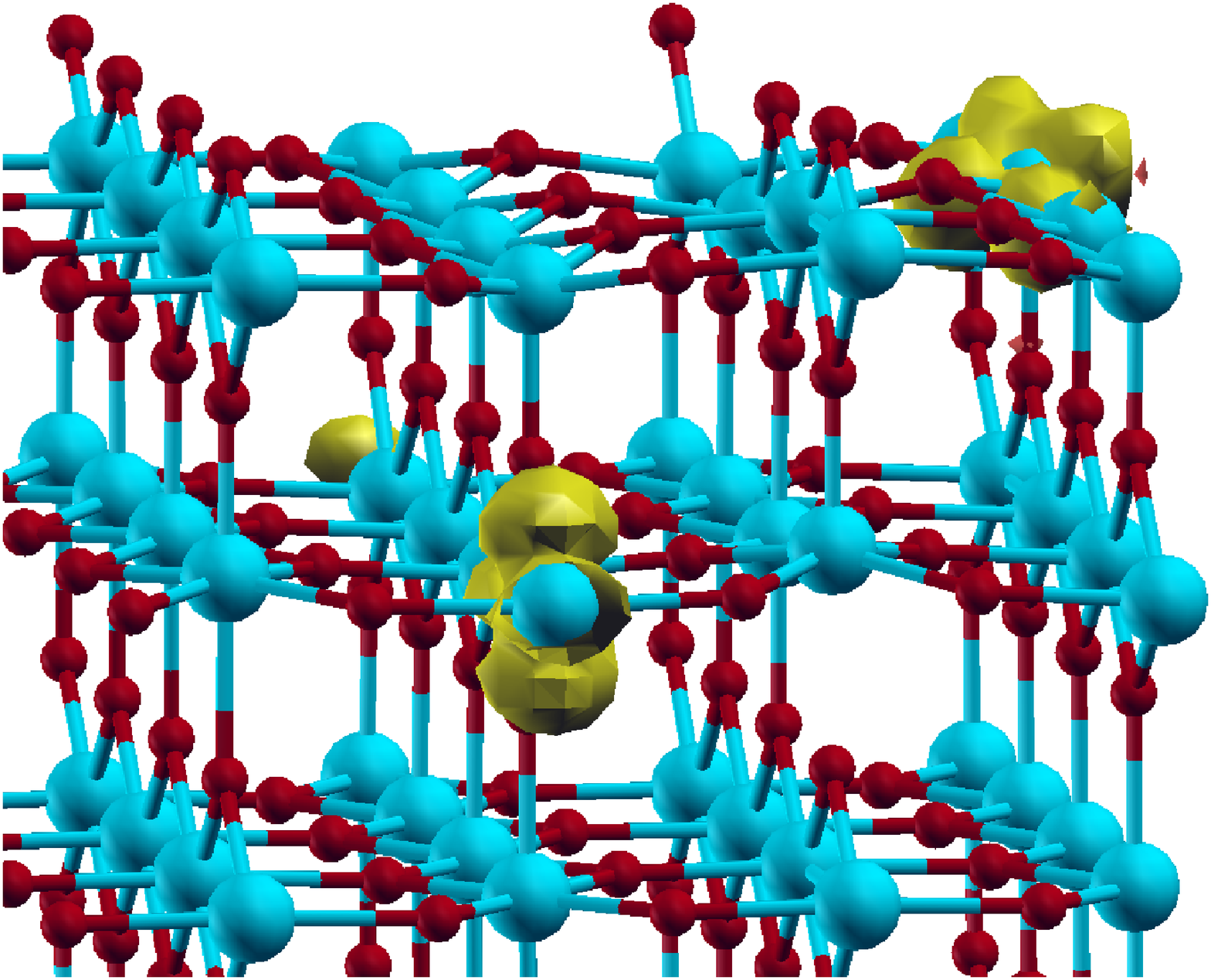}} \\
     \end{tabular}
    \caption{%
(a) Ball and stick model of the defective $\mathrm{TiO_{2}(110)}$ surface.
Red and blue spheres are $\mathrm{O}$ and $\mathrm{Ti}$ atoms, respectively. 
Panels (b), (c), and (d) depict the spin density (at 0.005~e/\AA$^{3}$) of three  
configurations from Table~\ref{tab:table1} with different charge localization topologies. 
\label{FIG0}
}
 \end{center}
\end{figure}

Using B3LYP on a c(4$\mathrm{\times}$2) slab with an O~vacancy,  
the defect charge is found to be localized on ${\it d}$-orbitals of  
two surface $\mathrm{Ti}$ atoms~\cite{divalentin}.
In particular, 
one unpaired electron is found on the under--coordinated $\mathrm{Ti(11)}$ site,  
while the other moves to an adjacent five--fold coordinated  
$\mathrm{Ti_{5c}}$ atom, such as $\mathrm{Ti(7)}$;
see Fig.~\ref{FIG0} for our site labeling scheme. 
By contrast, $\mathrm{LDA/GGA+U}$ studies~\cite{thornton,pacchioni,morgan,morgan1,filippone,CHS08} 
on the reduced $\mathrm{TiO_{2}(110)}$ surface have reported charge localization  
on different surface and/or subsurface sites. 
For instance, 
a combination of surface and subsurface localization immediately beneath the defect
on $\mathrm{Ti(11)}$ and $\mathrm{Ti(27)}$ (see Fig.~\ref{FIG0}) has been found~\cite{thornton}.
However, the results are reported to be strongly dependent on the  
supercell
size.
Using a (2$\mathrm{\times}$1) cell, the electrons are found at
$\mathrm{Ti(7)}$ and $\mathrm{Ti(23)}$, while using a
(4$\mathrm{\times}$1) cell complete subsurface localization is observed~\cite{CHS08}
at $\mathrm{Ti(23)}$ and $\mathrm{Ti(39)}$. 
On the other hand, a (4$\mathrm{\times}$2) cell yields
localization of the electrons on $\mathrm{Ti(11)}$ and $\mathrm{Ti(12)}$  
when using $U \geq 4.2$~eV, whereas smaller values of $U$ lead instead to delocalization~\cite{morgan1}.
Interestingly, 
some recent $\mathrm{GGA+U}$ studies~\cite{deskins2,dupuis} which focussed on the intrinsic  
electron transport in titania provide useful hints:
Electron hopping in defect--free
$\mathrm{TiO_{2}}$ bulk~\cite{deskins2} is described by a 
polaron localized at a $\mathrm{Ti^{3+}}$ site, which can hop to  
an adjacent $\mathrm{Ti^{4+}}$ with activation energies of only $\approx 0.1$~eV.
Furthermore, static calculations of ideal $\mathrm{TiO_{2}(110)}$ surfaces 
with a single excess electron suggest that the most favorable trapping sites  
are subsurface
ions that lie below rows of $\mathrm{Ti_{5c}}$ atoms on the surface~\cite{dupuis}.

In summary, the existing results of electronic structure calculations which suggest
highly localized excess electrons at very specific $\mathrm{Ti}$ sites  
are at odds with the current experimental picture of a more ``fuzzy'' scenario
involving many sites and delocalized excess charge.
However, low activation energies (of the order of 0.1~eV) and the multitude
of topologies for the localization of excess charge that have been previously revealed suggest
that the energy landscape might be relatively flat and thus prone to thermal  
fluctuations which could affect localization scenarios in a dynamical sense.

In this Letter, we describe 
extensive $\mathrm{GGA+U}$ {\it ab initio}  
molecular dynamics (AIMD) simulations~\cite{marx-hutter-book} 
of the excess charge (de--)localization dynamics as  
induced by temperature.
We find
that the defect charge 
created by an O~vacancy on $\mathrm{TiO_{2}(110)}$
is {\em dynamically shared} by
several subsurface and surface $\mathrm{Ti}$ sites, with a dominant contribution from
particular second layer subsurface sites which do not belong to $\mathrm{O_{\rm b}}$ rows.  
Thermal fluctuations of the titania 
lattice allow the excess charge to probe~-- and thus to populate~-- 
various local (electronic structure) minima of similar
energy but vastly different localization topologies. 
Our findings 
demonstrate the need to go beyond static optimization 
in order to uncover the dynamical nature of the phenomenon.
These results strongly support the proposal of a delocalized nature 
of the excess charge for such defects,
and in addition, 
they also reconcile the hitherto contradictory viewpoints 
from different static electronic structure calculations.

The reduced $\mathrm{TiO_{2}(110)}$ surfaces have been modeled by four
$\mathrm{O}$--$\mathrm{Ti}_2\mathrm{O}_2$--$\mathrm{O}$ tri--layer (4$\mathrm{\times}$2)
supercell slabs, separated by more than 10~{\AA}
(see Fig.\ref{FIG0}). 
The bottom of the slabs was passivated with pseudohydrogen
atoms of nuclear charge $+4/3$ and $+2/3$ in order to achieve well
converged results, which is our ``standard setup''~\cite{piotr}.
All calculations have been performed using spin--polarized $\mathrm{GGA+U}$,
PBE~\cite{pbe},
and ultrasoft pseudopotentials~\cite{vanderbilt}. 
Extending our previous work~\cite{piotr}, we have implemented
$\mathrm{GGA+U}$ into {\tt CPMD}~\cite{cpmd} 
using the self--consistent linear response  
approach~\cite{CG05,KC06} to compute the Hubbard
parameter, yielding $U = 4.2$~eV for our setup; 
{%
as usual the localization details will depend
on the particular value of
$U$, but the qualitative features reported in this
\mbox{paper} were checked
to be stable upon reasonable variation. 
}%
The occupations of the ${\it d}$-orbitals are calculated using
atomic--like wavefunction projectors. 
The lowest tri--layer atoms are constrained to their equilibrium
positions, while all other atoms are free to move.
The AIMD simulations~\cite{marx-hutter-book} 
were on the order of 10~ps in length, and used
the Car--Parrinello propagation technique~\cite{car}
with a fictitious electron 
mass of 700~a.u. and a time step of 0.145~fs.

In order to 
reveal 
the nature and distribution of the defect charge,
we performed
the $\mathrm{PBE+U}$ simulations at various temperatures, ranging from 700 to 1000~K.
This is well below the melting point of almost 2000~K,
but sufficiently far above the ambient temperature 
that the phonon dynamics is accelerated and  
thus the sampling of the potential energy surface (PES) is enhanced
on the picosecond AIMD time scale.
Two complementary scenarios,
characterized 
by different defect charge localization topologies,
were employed in order to provide two distinct sets of
initial conditions.
One 
scenario 
represents a case where the two excess
electrons 
are trapped at two second layer sites,
$\mathrm{Ti(24)}$ and $\mathrm{Ti(30)}$, under $\mathrm{Ti_{5c}}$
rows. 
The
second
scenario 
corresponds to localization on an under--coordinated site
in the first layer, $\mathrm{Ti(11)}$, and on $\mathrm{Ti(29)}$ in the second layer
under a $\mathrm{Ti_{5c}}$ row. 
In both cases the paramagnetic triplet state is 
preferred
and the
structure with defect charge localized on the second layer is 
roughly 0.6~eV
more stable (see Table~\ref{tab:table1}).

\begin{table}
\caption{\label{tab:table1} 
  Relative energies (in eV) and Ti(11)--O bond
  lengths (in {\AA}) of reduced $\mathrm{TiO_{2}(110)}$ 
  with different localization of the $\mathrm{O_{\rm v}}$--induced excess
  charge. The last three columns show on which layer(s) the two excess
  electrons are localized. Bold labels refer to the configurations in Fig.~\ref{FIG0}.
}
\begin{ruledtabular}
\begin{tabular}{cccccccc}
   &  $\mathrm{Ti^{3+}}$ & $\mathrm{Ti^{3+}}$&
 $\mathrm{\Delta E}$ & Ti(11)--O & 1$^{\rm st}$--layer& 2$^{\rm nd}$--layer & 3$^{\rm rd}$--layer \\
\hline
&$\mathrm{Ti(6)}$     &$\mathrm{Ti(32)}$  & 0.23 & 1.87 & 1 e & 1 e
 & 0 e\\
&{\bf Ti(15)}    &{\bf Ti(21)}  & 0.28 & 1.87 & 1 e & 1 e
 & 0 e\\
&$\mathrm{Ti(11)}$    &$\mathrm{Ti(29)}$  & 0.77 & 1.92 & 1 e & 1 e
 & 0 e\\
&{\bf Ti(24)}    &{\bf Ti(31)}  & 0.00 & 1.87 & 0 e & 2 e
 & 0 e\\
&$\mathrm{Ti(22)}$    &$\mathrm{Ti(29)}$  & 0.04 & 1.87 & 0 e & 2 e
 & 0 e\\
&$\mathrm{Ti(24)}$    &$\mathrm{Ti(30)}$  & 0.16 & 1.86 & 0 e & 2 e
 & 0 e \\
&$\mathrm{Ti(27)}$    &$\mathrm{Ti(29)}$  & 0.22 & 1.86 & 0 e & 2 e
 & 0 e \\
& {\bf Ti(30)}    & {\bf Ti(40)}  & 0.36 & 1.87 & 0 e & 1 e
 & 1 e\\
\end{tabular}
\end{ruledtabular}
\end{table}

\begin{figure}
  \begin{center}
    \begin{tabular}{cc}
      \subfigure[]{\includegraphics[scale=0.18,angle=-90]{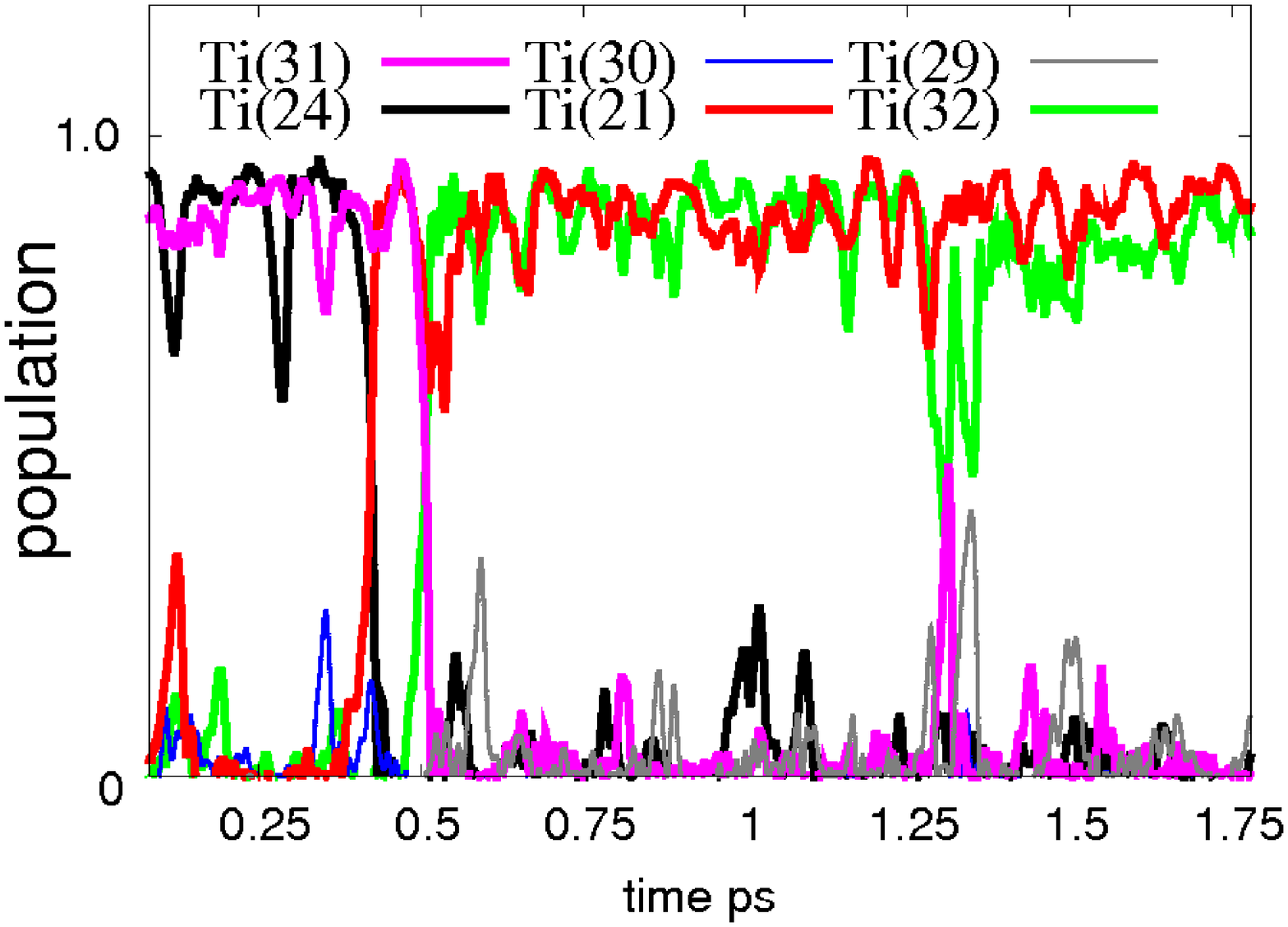}} &
      \subfigure[]{\includegraphics[scale=0.18,angle=-90]{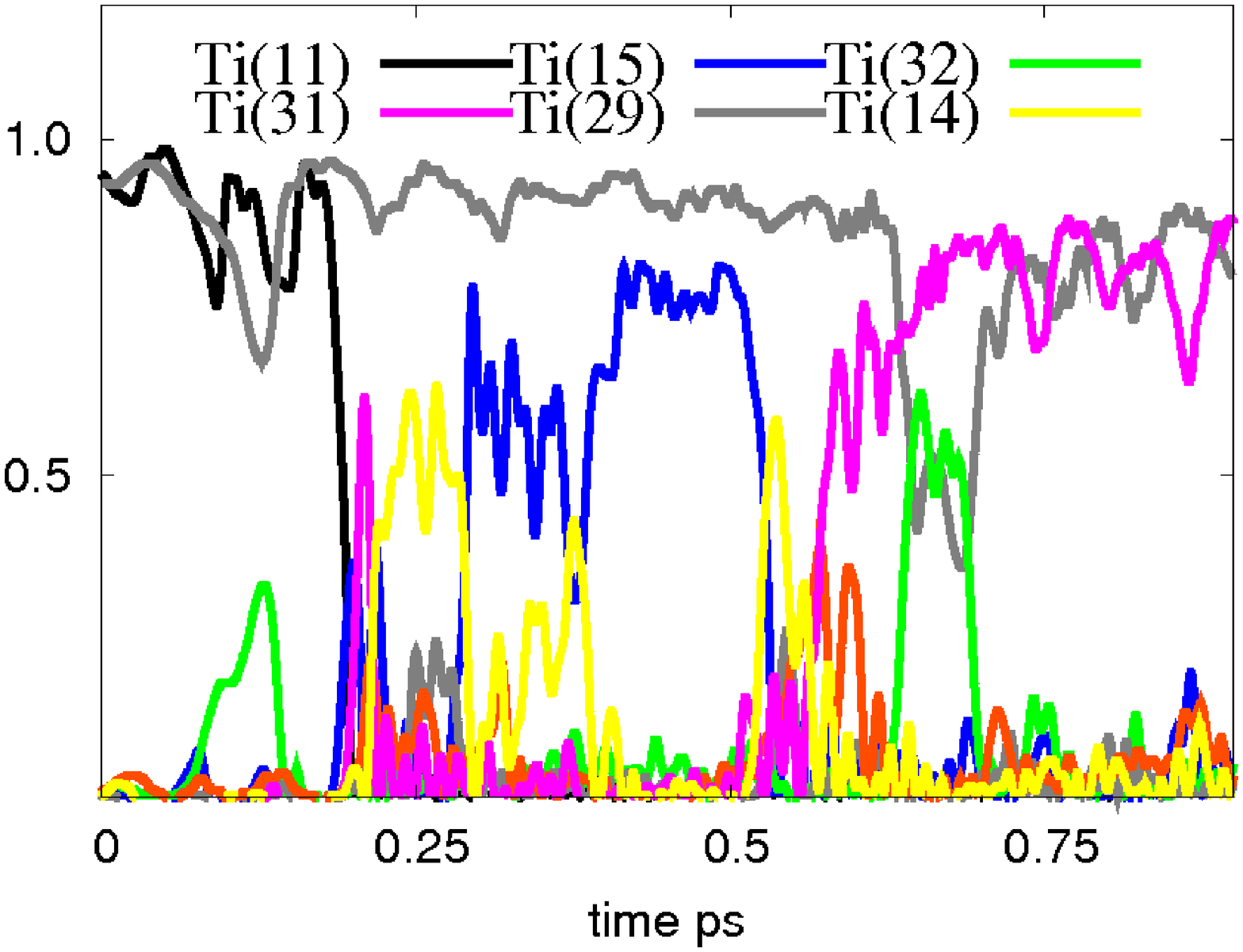}} \\[-4pt]
      \subfigure[]{\includegraphics[scale=0.18,angle=-90]{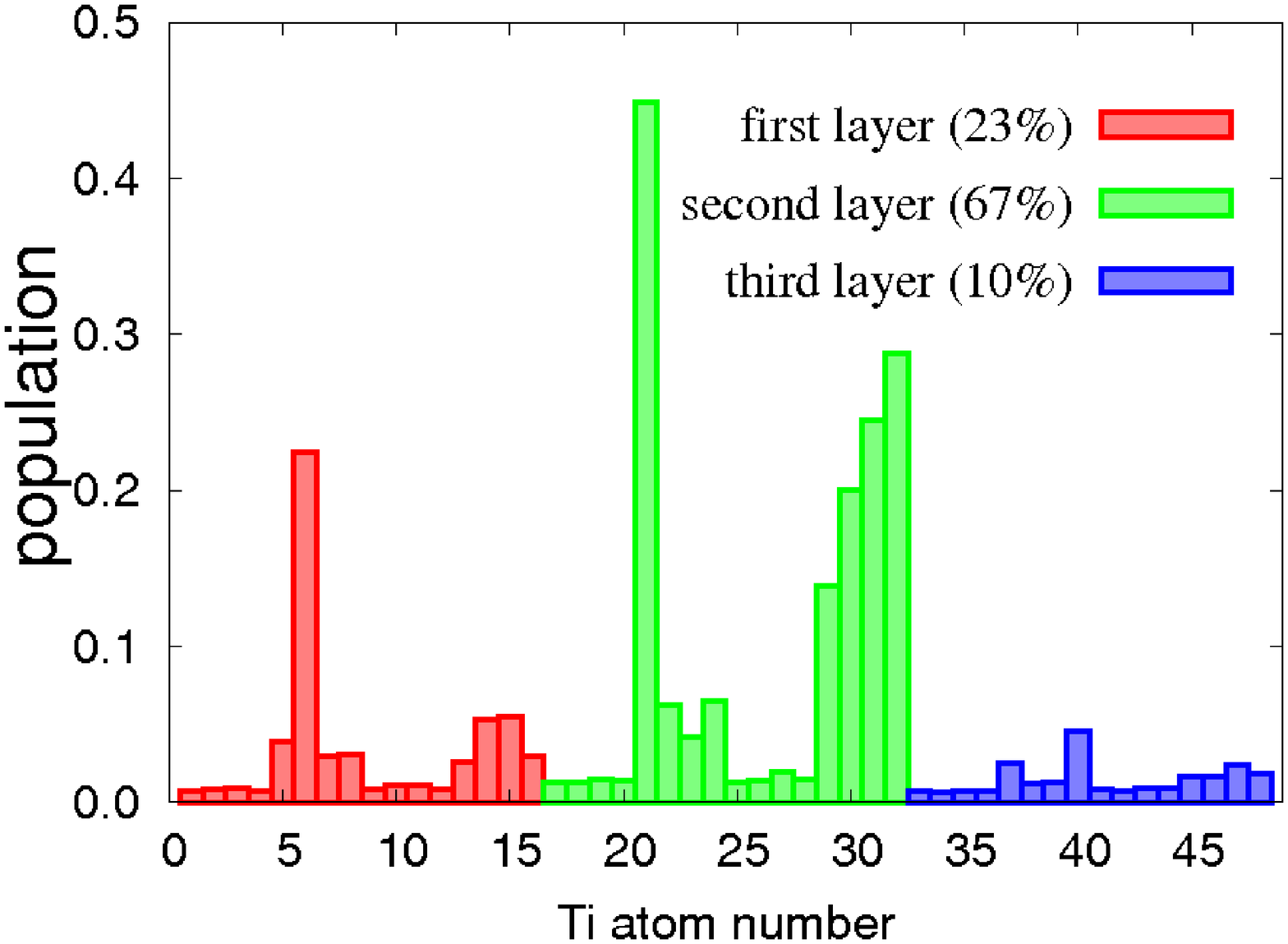}} &
      \subfigure[]{\includegraphics[scale=0.185,angle=-90]{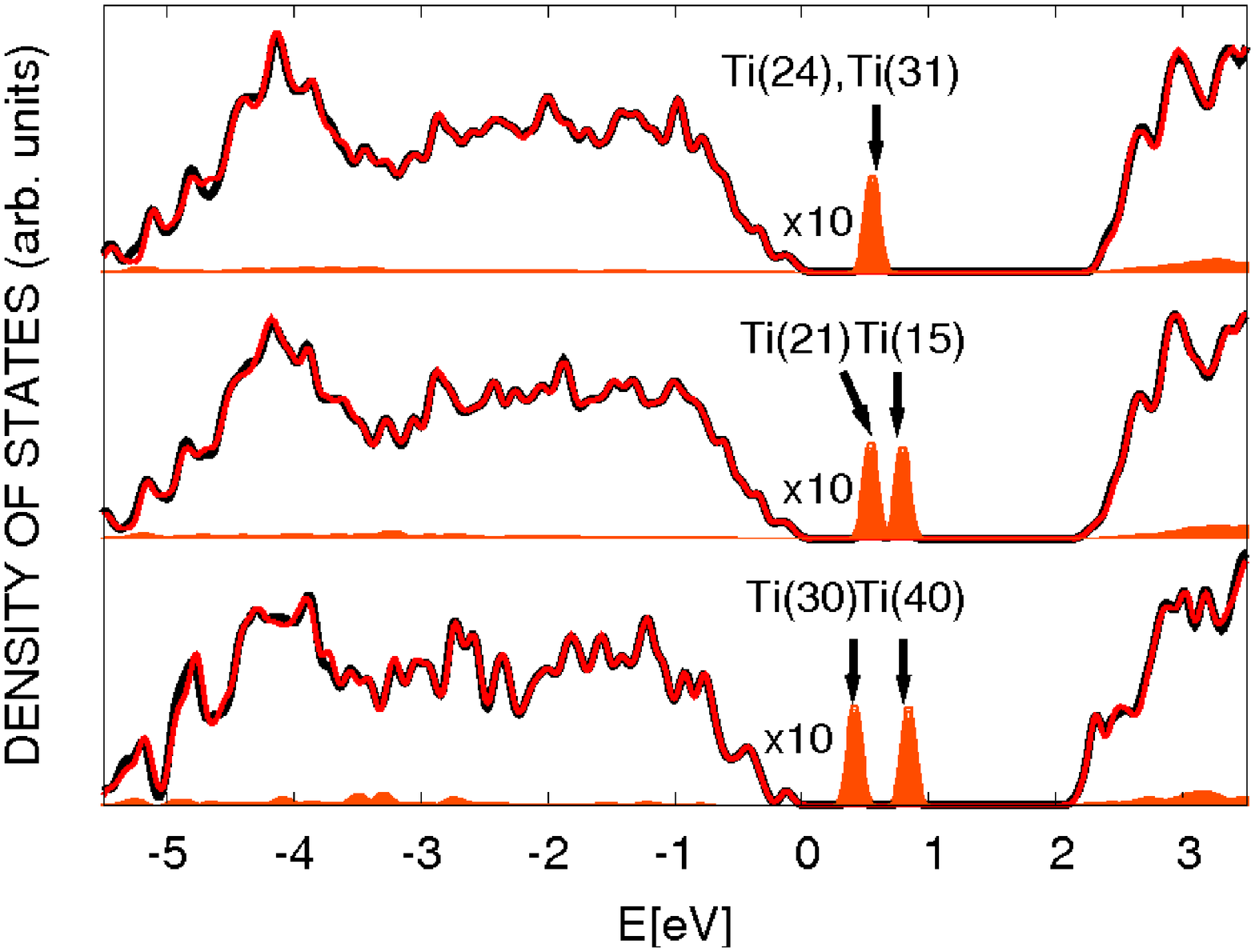}}
    \end{tabular}
    \caption{%
(a) 
Dynamics of the fractional occupation of particular $\mathrm{Ti}$ ${\it d}$--orbitals 
during a time fragment of the $\mathrm{PBE+U}$ simulation
carried out at 1000~K;
at $t=0$ the charge is localized in the
second subsurface layer at $\mathrm{Ti(24)}$ and $\mathrm{Ti(31)}$;
populations of
about $1$ and $0$ correspond to $\mathrm{Ti^{3+}}$
and $\mathrm{Ti^{4+}}$
charge states, respectively.
(b) 
Same as (a) but the charge is initially localized in the first layer 
at $\mathrm{Ti(11)}$ and subsurface at $\mathrm{Ti(29)}$ at 700~K.
(c) 
Distribution function of the average population of all available Ti--sites 
by the two excess electrons obtained from the full simulation underlying (a);
inset provides layer--averaged populations. 
(d) 
Electronic DOS ($\mathrm{\alpha}$/$\mathrm{\beta}$ spins: red/black lines)
of the three structures from Fig.~\ref{FIG0};
peaks in the band gap correspond to reduced Ti--sites;
the top of the valence band is set to 0~eV,
and the red area filling the gap states
(scaled by 10 to enhance visibility)
stems from $\mathrm{Ti^{3+}}$ sites as labeled.
}
\label{FIG1}
 \end{center}
 \end{figure}

After equilibrating both setups at 300~K for about 2~ps  
each, the systems were heated to their respective simulation temperatures.
The charge localization dynamics was monitored by computing, as a function of time, the
occupation matrix of each $\mathrm{Ti}$ ${\it d}$--$\mathrm{\alpha}$
and ${\it d}$--$\mathrm{\beta}$
spin orbital along the trajectories.
We will first consider the effect induced by thermal fluctuations  
on the setup with all excess charge initially
localized in the second layer. 
As demonstrated by Fig.~\ref{FIG1}(a), the defect charge  
transfers from sites
$\mathrm{Ti(24)}$ and $\mathrm{Ti(31)}$ to $\mathrm{Ti(21)}$ and $\mathrm{Ti(32)}$
in less than 0.5~ps;
{%
we note in passing that a qualitatively similar scenario
is observed using
a smaller (2$\mathrm{\times}$1) cell.
}%
Approximately 30 such charge migration events were observed during the 6~ps simulation,
although only a few of them involve charge leaving the second subsurface layer and moving to the first or third layer.
This implies that the  
defect charge
spends most of the time at second--layer
$\mathrm{Ti}$ ions,
but translocates
very rapidly from a given $\mathrm{Ti}$ site to an adjacent one within the
same subsurface row.
We observed no charge transfer to $\mathrm{Ti}$
sites belonging to adjacent rows.
Other interesting 
electronic topologies explored by the dynamics are cases in which the two
excess electrons are localized simultaneously on second-- and first--layer
sites, for example, on $\mathrm{Ti(32)}$ and $\mathrm{Ti(6)}$ or on
$\mathrm{Ti(21)}$ and $\mathrm{Ti(15)}$.
Furthermore, a configuration has been sampled where 
the excess charge is shared between the second and third layer involving 
$\mathrm{Ti(30)}$ and $\mathrm{Ti(40)}$.
The average lifetime for a specific charge--localized topology is roughly 0.3~ps
at 1000~K, 
which underscores the pronounced dynamical nature of the phenomenon. 
%
{%
At 700~K, this timescale increases to
about $0.4$~ps, from which
naive Arrhenius extrapolation yields 
a formal activation energy of approximately 0.07~eV and 
lifetimes of the order of ps and 
ns to $\mu$s 
at ambient and liquid nitrogen tem\-per\-atu\-res, respectively.
This implies that dynamical averaging up to quenched disorder 
at liquid helium conditions might be operational in surface science experiments.
}%

Next 
we consider
the complementary scenario where 
the 
defect charge
is localized initially 
in both the first and second layers at $\mathrm{Ti(11)}$ and $\mathrm{Ti(29)}$
(see Fig.~\ref{FIG1}(b)).  
After only 0.1--0.2~ps, the charge at $\mathrm{Ti(11)}$ transfers
from the $\mathrm{O_{b}}$ surface row to the adjacent surface row of
$\mathrm{Ti_{5c}}$ atoms where it is delocalized over two nearest--neighbor
sites $\mathrm{Ti(14)}$ and $\mathrm{Ti(15)}$. 
It remains
shared between these two sites for about 0.5~ps before it
jumps to the nearest--neighbor second--layer
atom, $\mathrm{Ti(31)}$. 
Here, 
charge transfer between the first and second layers
is found to be mediated by a charge delocalization process involving 
first--layer $\mathrm{Ti}$ sites in the $\mathrm{Ti_{5c}}$ row.  
Once the charge reaches the second layer, during the next
$\approx 4$~ps it visits 
the same $\mathrm{Ti}$ sites as it had earlier explored.
In one of these events a flipping of one of the two unpaired electrons between
an $\alpha$-- and $\beta$--${\it d}$ orbital occurs along the trajectory.

The dynamical behavior can be cast into distribution functions 
using localization histograms, which is illustrated 
in Fig.~\ref{FIG1}(c) for the dynamics in (a).
In both set\-ups 
the excess charge is found to be extremly mobile and to eventually
visit all available Ti~sites,
although it displays a strong preference for populating sites in  
the second subsurface layer 
under the $\mathrm{Ti_{5c}}$ rows (roughly 70~{\%}).
This is followed by population of surface states
($\approx 20$~{\%}),
and localization in the third subsurface layer ($\approx 10$~{\%}).
We note that in all observed charge transfer events
there was always a mediating nearest--neighbor $\mathrm{Ti}$ site
transiently involved in the process.
This dynamical scenario is in substantial agreement with the viewpoint 
suggested based on recent experiments~\cite{morgante,minato}.

In order 
to assess the energetics of the various localization topologies,
we have characterized representative structures in detail.
Six additional 
excess charge topologies
were obtained by quenching a large set of sampled configurations  
to $T=0$~K
using Car--Parrinello annealing~\cite{marx-hutter-book}
followed by standard optimization (see Table~\ref{tab:table1}).
Localization of one excess electron 
at $\mathrm{Ti(11)}$
induces a significant elongation of the  $\mathrm{Ti^{3+}(11)}$--$\mathrm{O}$
distance (see Table~\ref{tab:table1}), which is as expected~\cite{divalentin,morgan,morgan1}.
The relative energies confirm that the most stable sites for 
charge localization are those where both electrons
are in the second subsurface layer 
under $\mathrm{Ti_{5c}}$ rows~-- in line with earlier findings~\cite{CHS08}
and consistent with our localization histograms.
About 
0.2--0.3~eV higher in energy
are the topologies where the excess charge is shared between surface
$\mathrm{Ti_{5c}}$
atoms
and second subsurface layer sites below $\mathrm{Ti_{5c}}$ rows. 
Topologies resulting in excess charge localization on 
both second and third subsurface layer sites under 
$\mathrm{Ti_{5c}}$ are less stable by 
about 0.3--0.4~eV.
In the least stable configuration,
(by 0.7--0.8~eV) 
one electron is at an under--coordinated surface site, $\mathrm{Ti(11)}$ at the
$\mathrm{O_{b}}$ vacancy, whereas the other electron has transferred to the second subsurface
layer under a $\mathrm{Ti_{5c}}$ row at $\mathrm{Ti(29)}$.
In one case, 
starting from a configuration in which the charge is 
delocalized over two surface
atoms, $\mathrm{Ti(14)}$ and $\mathrm{Ti(15)}$,
and one second layer site, $\mathrm{Ti(29)}$, we end up
with a solution in which one electron is trapped 
at $\mathrm{Ti(29)}$ under the $\mathrm{Ti_{5c}}$ row 
and the other electron is localized on the $\mathrm{Ti}$ site  
directly below the $\mathrm{O_{b}}$ vacancy.  
This configuration, which has also been recently reported~\cite{thornton},
is found to be 0.22~eV higher
in energy than our lowest energy configuration.  
{%
The concomitant static distortions obtained from the nearest--neighbor Ti$^{3+}$--O
distances of the optimized structures
(b)--(d) in Fig.~\ref{FIG0}
relative to the non--relaxed reference~(a) 
are typically 0.05--0.1~{\AA}.
However, already at 300~K the thermal fluctuations are found to override
these static distortions by inducing larger 
root--mean--square--deviations of the same relative distances. 
}%

{%
Combining our static and dynamic calculations demonstrates that
the excess charge dynamics is intimately tied to the existence
of many close--lying minima on the
PES
which can be populated at finite temperatures~\cite{selloni}.
}%
Charge localization pattern from static optimizations therefore depend
heavily on the initial configuration due to trapping in local energy minima.
This might contribute to the variety of differing results
reported in the literature.
In this situation, AIMD helps greatly by not only providing a dynamic
perspective, but also by being able to explore vastly different localization
topologies.

Finally, the total and projected electronic 
density of states (DOS) of 
the distinctly different                                                                  
charge localization scenarios are analyzed in Fig.~\ref{FIG1}(d).  
In all 
cases, the O~vacancy                                                                    
leads to the appearance of distinct, filled states                                     
in the $\mathrm{TiO_{2}}(110)$ band gap, whose projection onto two specific              
$\mathrm{Ti}$ sites~-- although they are different for each 
of the three scenarios~-- fully accounts for the entire peak integral area.
Together with the almost instantaneous changes in localization patterns
embodied in Fig.~\ref{FIG1}(a) and (b), 
this indicates that the two excess electrons are 
essentially always
trapped at two well--identified, specific $\mathrm{Ti}$ sites.
These sites, however, are not unique for a given defect,
but rather they interchange dynamically.

In conclusion, 
finite temperature $\mathrm{PBE+U}$ simulations paint a complex
picture of the dynamics of electrons in defects at the reduced
titania TiO$_2$(110) surface. 
The excess charge, being trapped at specific $\mathrm{Ti}$ sites, 
migrates easily by phonon-assisted (thermally activated)  
hopping to other $\mathrm{Ti}$ sites, 
thus exploring significantly different electronic structure topologies.  
Due to averaging effects, this leads
to an ``effective delocalization'', which is preferentially found
on second--layer subsurface $\mathrm{Ti}$
atoms underneath
$\mathrm{Ti_{5c}}$ rows. 
Although much less frequently, excess charge also ``visits'' first layer 
and  
third layer subsurface sites.
We expect that this defect--induced complex charge \mbox{(de--)localization}  
dynamics scenario is not only fundamental to titania, but of broad significance
to oxide materials in general.

\acknowledgments
This work has been supported by the German Research Foundation (DFG) via
SFB~558, 
by RD~IFSC,
and by FCI.
Computational 
resources were provided by {\sc Bovilab@RUB} (Bochum) as well as RV--NRW.


\end{document}